\documentclass[10pt,letterpaper]{article}

\usepackage{cogsci}
\cogscifinalcopy 
\usepackage{pslatex}
\usepackage{apacite}
\usepackage{float} 
\usepackage{graphicx}

\title{Modeling Reliance on XAI Indicating Its Purpose and Attention}

\author{{\large \bf Akihiro Maehigashi (maehigashi.akihiro@shizuoka.ac.jp)} \\
  Shizuoka University, Shizuoka, Japan \\
\AND {\large \bf Yosuke Fukuchi (fukuchi@nii.ac.jp)} \\
National Institute of Informatics, Tokyo, Japan \\
\AND {\large \bf Seiji Yamada (seiji@nii.ac.jp)} \\
National Institute of Informatics, Tokyo, Japan\\ The Graduate University for Advanced Studies, SOKENDAI, Tokyo, Japan}
  
\begin{document}

\maketitle
\begin{abstract}
This study used XAI, which shows its purposes and attention as explanations of its process, and investigated how these explanations affect human trust in and use of AI. In this study, we generated heat maps indicating AI attention, conducted Experiment 1 to confirm the validity of the interpretability of the heat maps, and conducted Experiment 2 to investigate the effects of the purpose and heat maps in terms of reliance (depending on AI) and compliance (accepting answers of AI). The results of structural equation modeling (SEM) analyses showed that (1) displaying the purpose of AI positively and negatively influenced trust depending on the types of AI usage, reliance or compliance, and task difficulty, (2) just displaying the heat maps negatively influenced trust in a more difficult task, and (3) the heat maps positively influenced trust according to their interpretability in a more difficult task.

\textbf{Keywords:} 
AI; XAI; Purpose; Attention; Heat map; Reliance; Compliance; Algorithm aversion; Trust; Structural equation modeling (SEM)
\end{abstract}

\section{Introduction}

In recent years, artificial intelligence (AI) has been entering all aspects of life. In the future, many human activities are expected to be performed with AI. Successful cooperation with AI requires human users to appropriately adjust their use of AI~\cite{Lee04}. A fundamental factor used to decide the levels at which to use AI is considered to be trust in AI~\cite{Lee04,Wiegmann01}. Trust is defined as ``the attitude that an agent will help achieve an individual's goals in a situation characterized by uncertainty and vulnerability''~\cite{Lee04}. Proper use of AI is achieved when trust is appropriately calibrated to its actual reliability and maximizes task performance. This process is called trust calibration.

Moreover, even though AI has apparently superior reliability, people avoid using it. This phenomenon is also known as algorithm aversion~\cite{Dietvorst2015algorithm}. Algorithm aversion is defined as ``a behavior of discounting algorithmic decisions with respect to one's own decisions or other's decisions.''~\cite{Mahmud2022what}. One of the factors causing algorithm aversion is considered to be the black-box nature of AI, that is, its lack of transparency~\cite{Glikson2020human,Mahmud2022what}. This problem is considered to be improved by providing explanations about AI algorithms, which would help people understand the underlying rationale behind AI performance~\cite{Kayande2009how}.

Recently, explainable AI (XAI) has been developed for making AI processes and outputs more understandable to humans by providing explanations about them~\cite{Gunning2019xaiexplainable}. XAI might help develop proper trust in AI and inhibit algorithm aversion. This study investigated how people develop trust and decide to use AI that shows its purposes and attention as explanations. 

\section{Related Work}
\subsection{Trust in AI and algorithm aversion}

Many studies have investigated trust in AI in accordance with its performance. Basically, people increase their trust and use of AI when AI shows high performance and decrease them when AI shows errors~\cite{Lee1992trust,Wiegmann01}. However, people are generally sensitive to AI errors and less tolerable of them, and therefore, even less severe AI errors cause extreme under-trust~\cite{Dzindolet02} and algorithm aversion~\cite{Dietvorst2015algorithm}. 

Useful ways of avoiding extreme under-trust and algorithm aversion because of AI errors include displaying reasons why AI causes certain errors~\cite{Dzindolet02}, showing how AI algorithms can work~\cite{Kayande2009how}, and providing user training on understanding AI algorithms and errors~\cite{Araujo2020ai,Bahner2008misuse}. These studies showed that making AI algorithms transparent by providing explanations about the algorithms improves trust calibration and inhibits algorithm aversion.

Moreover, human emotion has been considered an important factor affecting trust in and use of AI. Emotional trust, such as feeling secure, comfortable, and content, has been investigated and distinguished from cognitive trust, that is, people's rational expectation that AI has the capability to perform well~\cite{Komiak2006effects,Zhang2021head}. Increasing the tangibility or anthropomorphism of AI could avoid an inappropriate decrease in emotional trust~\cite{Glikson2020human}.

\subsection{XAI}
In recent years, various studies have focused on XAI. While the structures of explanations have variations including decision-trees, rules, classifiers, and saliency maps, different algorithms have been developed for each structured explanation.~\cite{Gunning2021darpas}. 

At the beginning of XAI research, LIME~\cite{Ribeiro2016why} effectively demonstrated the importance of XAI and presented concrete XAI algorithms for classification learning. Since then, a major approach for XAI has been to indicate important features, including SHAP~\cite{Lundberg2017unified} and influence functions~\cite{Koh2017understanding}, and to reevaluate the transparency of machine learning algorithms~\cite{Rudin2019stop}. 

The most popular approach in XAI for CNN-based image recognition is Grad-CAM~\cite{Selvaraju2017grad} to analyze CNN models and generate heat maps indicating important features (attention) in an original image. This approach is also called attention-based XAI. However, this XAI has disadvantages in that humans have much difficulty interpreting how AI recognizes an image with only heat maps.

To overcome the limitation of attention-based XAI, novel studies on XAI done to focus on human cognitive processes, including purposes and interpretation, are beginning because the final target of XAI is to generate good explanations with which humans can understand an AI well. CX-ToM~\cite{Akula2022cx} provides dialogue-based explanations instead of heat maps to explain CNN learning models. Theory of mind (ToM) is a key concept for generating the explanations. 

Sanneman and Shah~\cite{Sanneman2022situation} proposed introducing situation awareness for designing and evaluating XAI systems. Their assertion is that XAI systems and explanations should be designed and evaluated significantly depending on the situation, task, and context.

\section{Hypotheses}

\begin{figure}[tbp]
    \begin{center}
    \includegraphics[width=70mm]{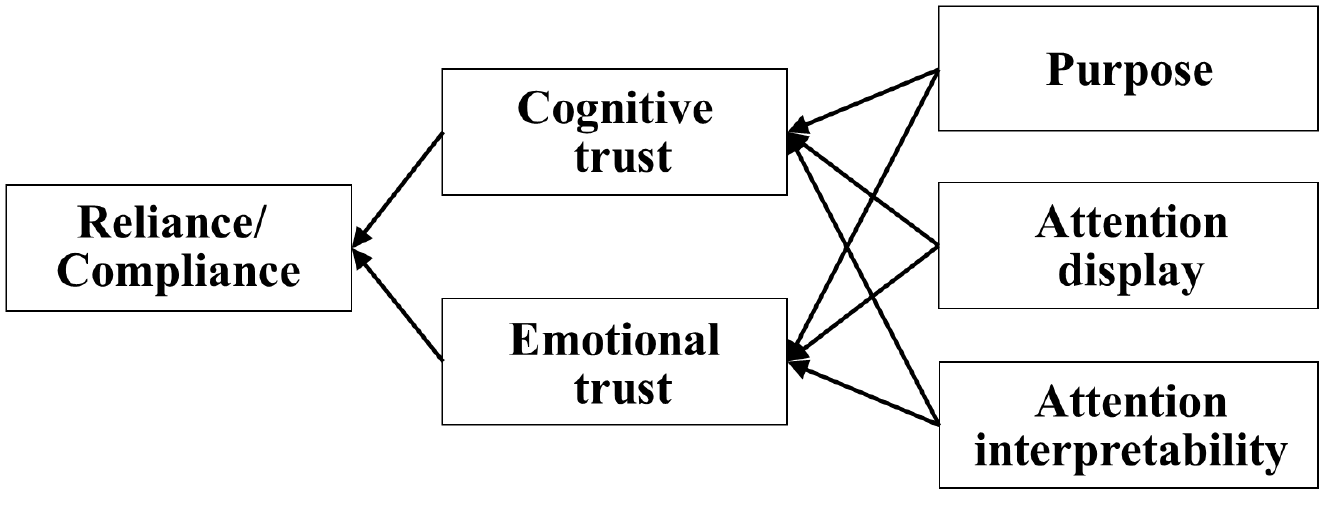}
    \end{center}
    \vspace{-4mm}
    \caption{Hypothetical model}\label{hypothesis-model}
\end{figure}

This study investigated how the purposes and attention of AI influence trust and use of AI. A hypothetical model is shown in Figure~\ref{hypothesis-model}. First, this study considers two different types of AI usage: reliance and compliance. In this study, \textbf{reliance} was defined as dependence on AI, such as delegating a task to an AI, and \textbf{compliance} was defined as acceptance of AI outputs, such as following the answers of an AI, as extended definitions from the previous studies~\cite{Kohn2021measurement,Vereschak2021how}.

Next, regarding the purpose of AI, humans' understanding of AI purposes plays an important role in developing trust in and using an AI~\cite{Akula2022cx}. Since ``goals'' or ``intentions'' are major components for ToM, the purposes of AI are expected to be a significant factor that increases cognitive and emotional trust \textbf{(H1)}.

Moreover, regarding AI attention, there are studies that found positive relationships between human interpretability of AI attention and trustworthy AI~\cite{Sanneman2022situation, Tomsett2020rapid}. Therefore, there is a possibility that just displaying AI attention would not influence cognitive and emotional trust \textbf{(H2)}; however, higher interpretability for AI attention is assumed to increase cognitive and emotional trust \textbf{(H3)}. The hypotheses are summarized as follows.

\begin{description}
    \item \textbf{H1: Displaying the purpose of AI positively influences cognitive and emotional trust.}
    \item \textbf{H2: Displaying AI attention does not influence cognitive and emotional trust.}
    \item \textbf{H3: The interpretability of AI attention positively influences cognitive and emotional trust.} 
\end{description}

In the following, first, we generated heat maps indicating AI attention. Next, we conducted Experiment 1 to confirm the validity of the interpretability of the heat maps. Finally, we conducted Experiment 2 to test the hypotheses using the validated heat maps.

\section{Heat Maps Indicating AI Attention}
First, we prepared two different types of image datasets, body-shape and human facial images, to conduct obesity screening and drowsiness detection in the following experiment. The body-shape images were used from a body-shape database~\cite{Moussally2016database}, and the human facial images were used from a drowsiness dataset~\cite{Ghoddoosian2019realistic}. We chose these experimental tasks because they were related to realistic healthcare problems. 

Next, we prepared deep-learning models to generate AI attention with Grad-CAM~\cite{Selvaraju2017grad} (Grad-CAM with PyTorch 17-05-18 https://github.com/kazuto1011/grad-cam-pytorch). A simple model composed of a four-layer convolutional neural network and two-layer perception was used for the obesity screening. Table~\ref{table:fatmodel} shows the structure of the model. Grad-CAM visualized the attention of the conv\_2 layer. The model for the drowsiness detection was Inception-ResNet~\cite{10.5555/3298023.3298188}, which is commonly used for face recognition~\cite{8373860,PENG20209}. Grad-CAM referred to the Inception-ResNet-B block of the model for attention visualization.

\begin{figure}[tbp]
    \begin{center}
    \includegraphics[height=60mm, width=73mm]{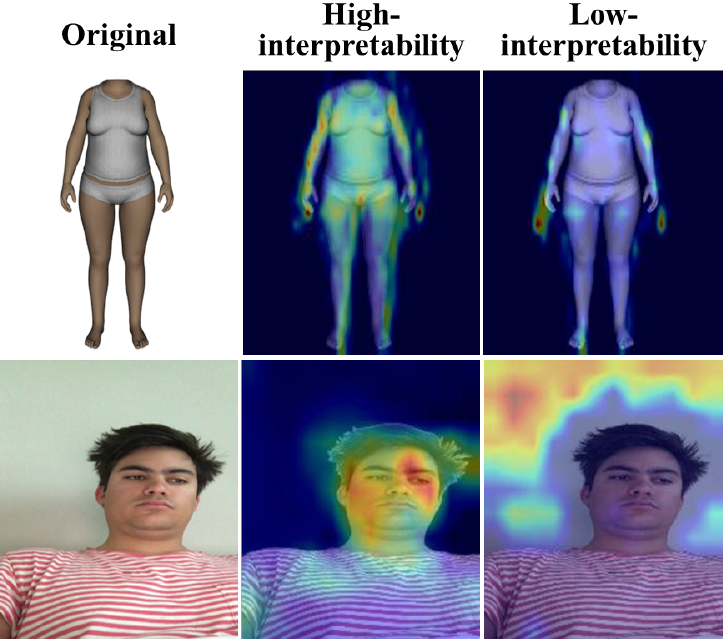}
    \end{center}
    \vspace{-4mm}
    \caption{Original images and heat maps with high and low interpretability. Top three images are for obesity screening, bottom three are for drowsiness detection.}\label{interpretability-image}
\end{figure}

Moreover, we developed high- and low-interpretability models for each task. To manipulate the interpretability of AI attention, we biased the training datasets and initial weights of the models. For the high-interpretability models, we removed the backgrounds of the images in the preprocessing because we found that the models overfit the dataset by focusing on marginal information, which was expected to lead to low interpretability for our tasks. In addition, the high-interpretability model for the drowsiness detection was initialized with parameters pretrained with VGGFace2~\cite{Cao18}, a dataset for face recognition, and trained in a transfer learning manner~\cite{tan2018survey}. The original images and the heat maps with high and low interpretability are shown in Figure~\ref{interpretability-image}.

\begin{table}[tbp]
  \caption{Structure of model for obesity screening task}\label{table:fatmodel}
  \small
  \renewcommand{\arraystretch}{1}

\begin{tabular}{lcl}
  \hline
  \textbf{Layer name} & \textbf{Output size} & \textbf{Parameters} \\ 
  \hline
Input & (3, 160, 160) &  \\ 
\hline
conv\_1 & (8, 77, 77) &  $7\times7$,  $stride=2$ \\
BatchNorm, ReLU, MaxPool & (8, 25, 25) & $3\times3$ \\ 
\hline
conv\_2 & (8, 23, 23) & $3\times3$, $stride=1$ \\
BatchNorm, ReLU, MaxPool & (8, 7, 7) & $3\times3$ \\ 
\hline
conv\_3 & (8, 5, 5) & $3\times3$, $stride=1$ \\
BatchNorm, ReLU, Flatten & (200) & \\ 
\hline
Linear, ReLU & (128) & \\ 
\hline
Linear, ReLU & (2) & \\ 
\hline
\end{tabular}
\end{table}

\section{Experiment 1}
\subsection{An online experiment system}
One-hundred heat maps created by each high- and low-interpretability model were selected by an experimenter for the obesity screening and drowsiness detection tasks. Each participant evaluated the 100 heat maps. To evaluate the maps, the GUI for the XAI in the online experiment system, in particular, the layout of the original image and the corresponding heat map, was designed on the basis of previous work~\cite{Lu2019deep, Rajaraman2020iteratively, Wehbe2021deepcovid}. Actually, the original image and the heat map were simply located side by side. 

\subsection{Method}
\subsubsection{Experimental design and participants}
The experiment had a two-factor between-participants design. The factors were the interpretability (high and low) and the task (obesity screening and drowsiness detection). A priori power analysis with G*Power indicated that 128 participants were needed for a medium effect size ($f = .25$) with the power at .80 and alpha at .05~\cite{Faul2007gpower}. A total of 200 participants were recruited through a cloud-sourcing service provided by Yahoo! Japan. They were randomly assigned to one of the conditions and conducted a task. However, 71 participants were detected as inattentive by the attention check items of the Directed Questions Scale (DQS)~\cite{Maniaci2014caring} and were excluded from the analysis. As a result, data of 129 participants (88 male and 41 female from 16 to 78 y/o, $M = 47.04$, $SD = 12.04$) were used.

\subsubsection{Procedure}

The participants first agreed with the informed consent and read the explanations of the experiment. In particular, the heat map was explained as indicating AI attention when the AI screened for obesity or detected drowsiness. After that, they evaluated 100 heat maps in accordance with their conditions. The order of the heat maps was randomized. During the evaluation, the participants were asked ``how much can you interpret how the AI screened for obesity (or detected drowsiness) based on the heat map of AI attention?'' and evaluated the interpretability on a 5-point scale (1: not interpretable at all - 5: extremely interpretable). Right after the evaluation, they were required to answer whether the body shape was normal or obese in the obesity screening or whether the person was awake or drowsy in the drowsiness detection task.

\subsection{Results and discussion}
\subsubsection{Interpretability score}
First, the mean interpretability score of each heat map was calculated. On the basis of the score, 50 high-scored heat maps in the high- and 50 low-scored heat maps in the low-interpretability conditions were selected for each task. Next, using the 50 heat maps in each condition, the mean interpretability score of each participant was calculated, and an ANOVA was performed on the score (Figure~\ref {result1-exp1}). There was a significant interaction $(F(1, 125)=6.13, p=.01, \eta_{p}^2=.05)$, and significant simple main effects showed that the score was higher for the high-interpretability condition than for the low one in the obesity screening $(F(1, 125) = 9.82, p < .01, \eta_{p}^2=.07)$ and in the drowsiness detection task $(F(1, 125) = 41.75, p < .01, \eta_{p}^2=.25)$. From these results, we confirmed 50 heat maps with validated high and low interpretability in both tasks.

\subsubsection{Task analysis}
\begin{figure}[tbp]
    \begin{center}
    \includegraphics[width=66mm]{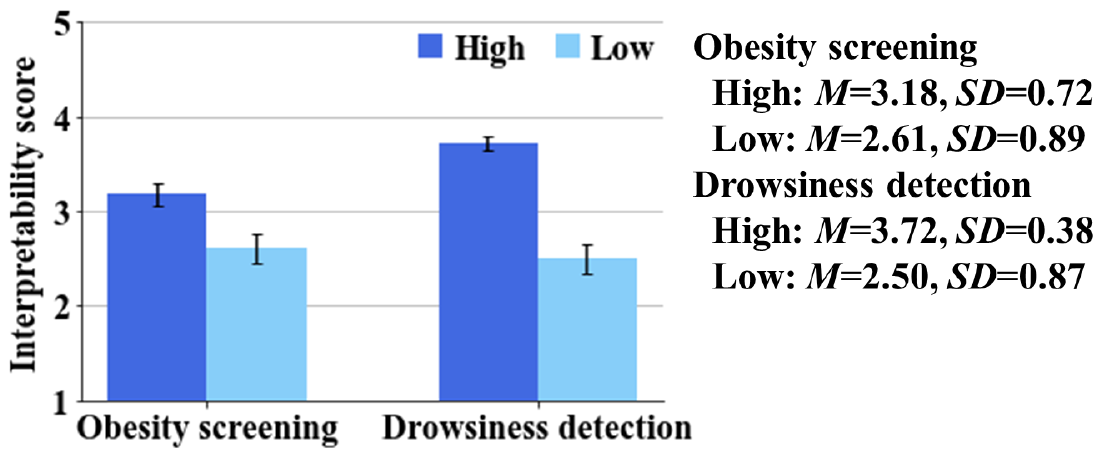}
    \end{center}
    \vspace{-4mm}
    
    \caption{Interpretability score for high-interpretability (High) and low-interpretability (Low) conditions in each task. Values show mean scores and standard deviations. Error bars show standard errors.}\label{result1-exp1}
    
    \begin{center}
    \includegraphics[width=85mm]{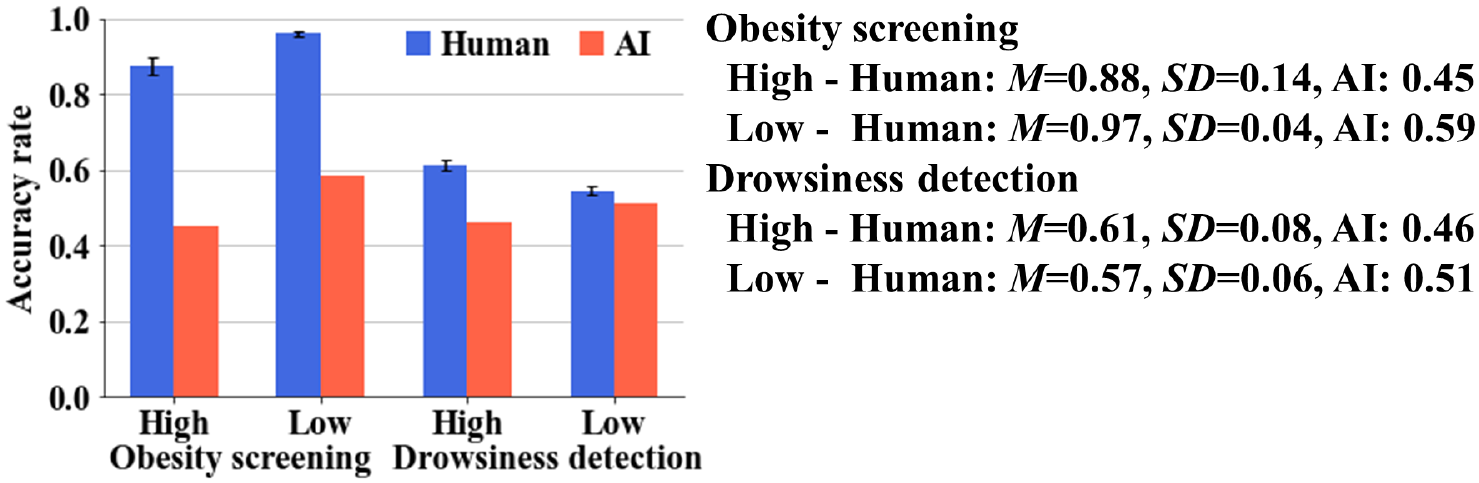}
    \end{center}
    \vspace{-4mm}
    \caption{Accuracy rate of participants (Human) and AI for high-interpretability (High) and low-interpretability (Low) conditions in each task. Values show mean human scores, standard deviations, and AI scores. Error bars show standard errors.}\label{result2-exp1}
\end{figure}

Using the 50 heat maps selected in each condition, the mean accuracy rate of each participant was calculated. Also, the mean of the accuracy rate of each AI model, indicated when the heat maps were created in the previous section, was calculated (Figure~\ref {result2-exp1}). First, as a result of an ANOVA on the mean accuracy rate of the participants in the four conditions, there was a significant main effect on the task factor showing that the rate was higher in the obesity screening than in the drowsiness detections task $(F(1, 125) = 431.17, p < .01, \eta_{p}^2=.78)$. Also, the results of one-sample t-tests showed that the rates of the participants were significantly greater than those of the AI in all the conditions $(ts>4.87, ps<.01, rs>.66)$. Statistical powers of higher than .80 were assured through post-hoc power analyses with G*Power. From these results, we confirmed that the obesity screening task was easier for the humans. Also, the humans performed better than the AI in all conditions.

\section{Experiment 2}
\subsection{Experimantal task}
This task was conducted using the original images and heat maps selected in Experiment 1. The experimental factors in this experiment were the usage-type (reliance and compliance), the purpose (with and without purpose), the attention-interpretability (high-, low-, and no-interpretability), and the task (obesity screening and drowsiness detection).

\begin{figure}[tbp]
    \begin{center}
    \includegraphics[height=70mm, width=85mm]{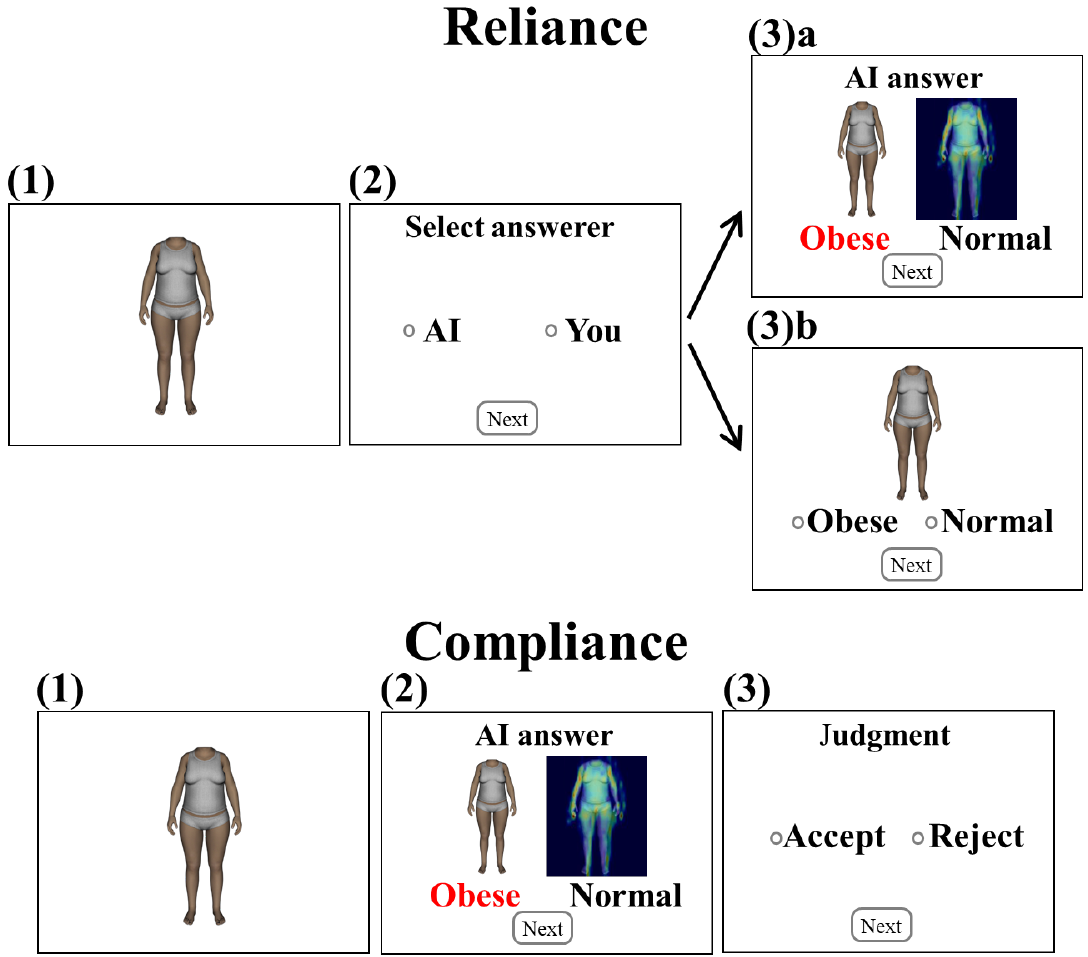}
    \end{center}
    \vspace{-4mm}
    \caption{Examples of task procedures for reliance and compliance. These examples are for high-interpretability condition without purpose in obesity screening task.}\label{task-paradigm}
\end{figure}

First, regarding the usage-type factor, the task procedures were set up according to the previous studies~\cite{Kohn2021measurement,Vereschak2021how} (Figure~\ref {task-paradigm}). The procedure for reliance was as follows: \textbf{(1)} An original image was displayed at the center of the display as a screening (or detection) problem for 5 seconds. \textbf{(2)} The participant decided to depend on the AI or themselves for the obesity screening (or drowsiness detection) by clicking. \textbf{(3)a} If the participant decided to depend on the AI, the AI showed its answer with a heat map (or without it). \textbf{(3)b} If the participant decided to depend on themselves, they answered whether the body shape was normal or obese in the obesity screening (or whether the person is awake or drowsy in the drowsiness detection) by clicking. 

Also, the procedure for compliance was as follows: \textbf{(1)} An original image was displayed at the center of the display as a screening (or detection) problem for 5 seconds. \textbf{(2)} The AI showed its answer with a heat map (or without it). \textbf{(3)} The participant decided to accept or reject the answer by clicking.

Moreover, regarding the purpose factor, in the with-purpose condition, the AI purpose stayed displayed on the bottom of the display during the task, stating ``the purpose of this AI is to screen for obesity based on body shape for healthcare counseling'' for the obesity screening and ``the purpose of this AI is to detect a state of fatigue based on expressions of drowsiness for healthcare counseling'' for the drowsiness detection. In comparison, in the without-purpose condition, there was no display of the purposes. 

Furthermore, regarding the attention-interpretability factor, the heat maps with high and low interpretability selected in Experiment 1 were respectively used for the high- and low-interpretability conditions in both tasks. The accuracy rates of the AI models were reflected in the answers during the task. Also, in the no-interpretability condition, there was no display of the heat maps, and the AI model for high or low interpretability was randomly assigned.

In addition, the participants were required to achieve as many correct answers as possible. Also, the partner AI was explained as learning from their answers and becoming smarter. This description was added to make participants feel that the AI was more realistic.

\subsection{Method}
\subsubsection{Experimental design and participants}
The experiment had a four-factor between-participants design. A total of 250 participants were recruited through a cloud-sourcing service provided by Yahoo! Japan. They were randomly assigned to one of the conditions and conducted a task. However, 20 participants were detected as inattentive by DQS and were excluded from the analysis. Also, 21 participants who showed exceptional behaviors, explained later, were excluded from the analysis. As a result, data of 209 participants (155 male and 54 female from 19 to 76 y/o, $M = 48.72$, $SD = 11.27$) were used.

\subsubsection{Procedure}
The participants first agreed with the informed consent and read the explanations about the task procedure. After that, they started the task. Each participant answered regarding the original image 50 times. The order of the images was randomized. During the task, after every 10 problems, they answered two types of trust questionnaires to measure cognitive and emotional trust. 

To measure cognitive trust, the Multi-Dimensional Measure of Trust (MDMT)~\cite{Ullman18} was used. MDMT was developed to measure a task partner's reliability and competence corresponding to the definition of cognitive trust. The participants rated how much the partner AI fit each word (reliable, predictable, dependable, consistent, competent, skilled, capable, and meticulous) on an 8-point scale (0: not at all - 7: very). Moreover, for emotional trust, we asked participants to answer how much the partner AI fit each word (secure, comfortable, and content) on a 7-point scale (1: strongly disagree - 7: strongly agree) as in the previous study~\cite{Komiak2006effects}.

In addition, in MDMT, participants could choose ``does not fit,'' which prevented possibly meaningless ratings. Twenty-one participants who chose it for all 8 words at the same time at least once during the task were considered to display exceptional behaviors and eliminated from the analysis. 

\subsection{Results}
\subsection{SEM analyses}
First, the data sets were prepared on the basis of the variables of the hypothetical model. For the usage-type (reliance and compliance) variable, the mean reliance or compliance rate was calculated every 10 problems. Next, for the purpose variable, the with- and without-purpose conditions were respectively represented using ``1'' and ``0'' as dummy variables. Also, for the attention-display variable, the high- and low-interpretability conditions were represented using ``1'' and the no-interpretability condition using ``0'' as dummy variables. Moreover, for the attention-interpretability variable, the interpretability score rated in Experiment 1 was used as a representative value of a heat map. The mean interpretability score was calculated in accordance with the displayed heat maps every 10 problems. Also, for the no-interpretability condition, ``0'' was used since there was no interpretability. Finally, the cognitive and emotional trust variables were treated as latent variables based on rated scores after every 10 problems.

\begin{figure*}[tbp]
    \begin{center}
    \includegraphics[width=160mm]{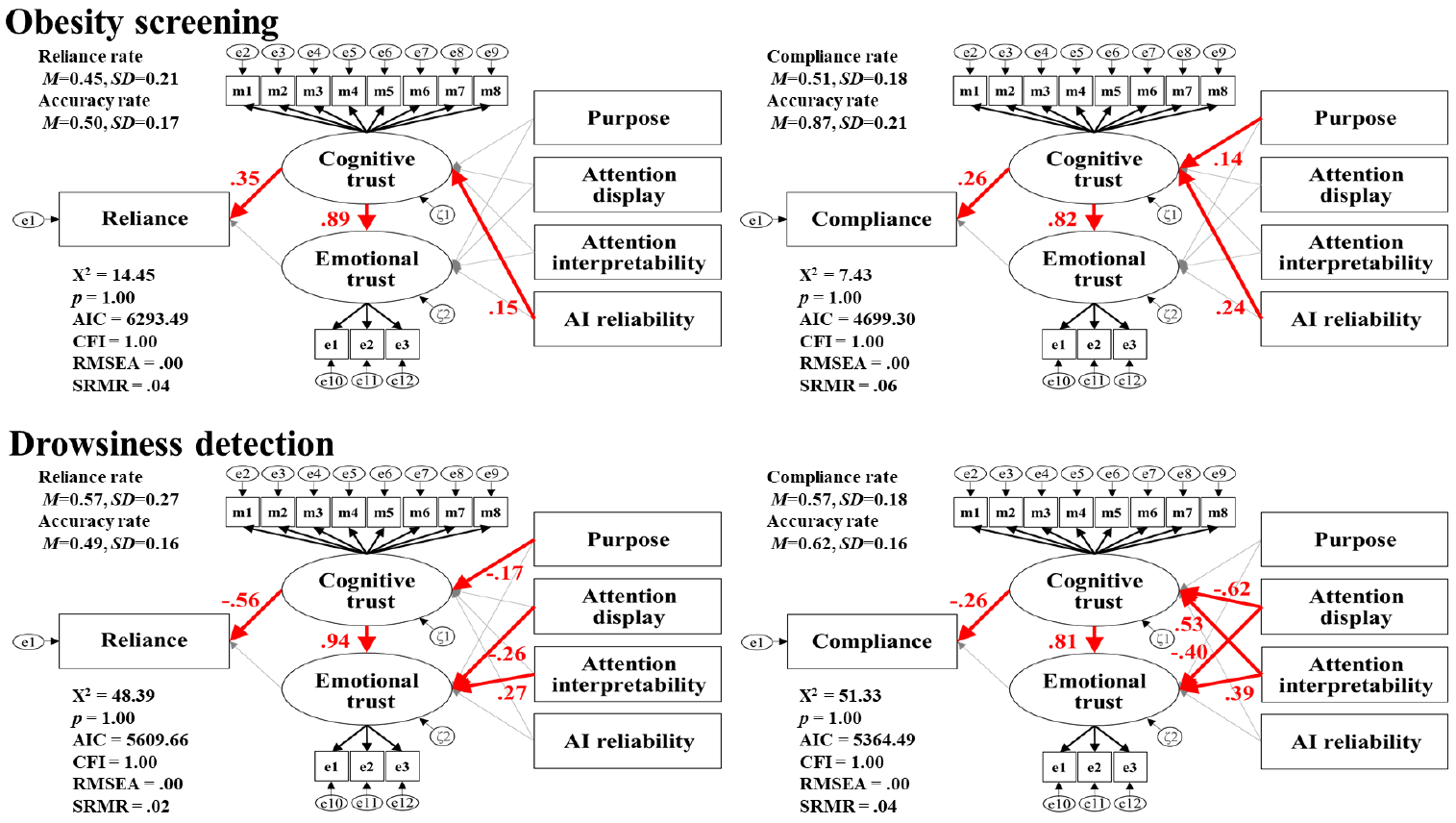}
    \end{center}
    \vspace{-4mm}
    \caption{Modified models and goodness-of-fit values. Values in red are standardized path coefficients, which were statistically significant. Regarding latent variables (cognitive and emotional trust), there were significant positive influences from latent variables on all observable variables (ratings in MDMT from m1 to m8 and emotional trust questionnaire from e1 to e3) for four models. Mean reliance or compliance rates, mean accuracy rates, and standard deviations in each task were also displayed.}\label{sem-models}
\end{figure*}

As a result, 5 data sets of each variable were created from each participant. However, there were 12 participants who rated ``does not fit'' at least once on the MDMT, so the corresponding data set was eliminated. Using the data sets, first, we developed reliance and compliance models for the obesity screening and drowsiness detection tasks on the basis of the hypothetical model. However, none of the models fit the data well. 

Next, we modified the models by making a path from cognitive to emotional trust as in the previous study~\cite{Komiak2006effects}. Also, we added the AI reliability variable, the mean accuracy rate of the AI every 10 problems, to make the path to cognitive and emotional trust. The modified models were fitted using robust maximum likelihood estimation~\cite{Rosseel2012lavaan}. As a result, all the models fit the data well, and the goodness-of-fit values met the criteria~\cite{Kline2011principles}. Figure~\ref{sem-models} shows the modified models and goodness-of-fit values. Statistical power analyses with R software with alpha at .05 revealed that the SEM performed with sample sizes ($N = 249$ and 274 for reliance and compliance models of the obesity screening, and 250 and 259 for reliance and compliance models of the drowsiness detection) for exact-fit tests obtained powers of .96, .98, .96, and .97, showing satisfactory statistical powers.

\subsection{Hypothesis survey}

Regarding \textbf{H1}, which is related to AI purpose, there was a positive influence from the AI purpose on cognitive trust for the compliance model in the obesity screening task. Also, there was a negative influence from the purpose on cognitive trust for the reliance model in the drowsiness detection task. Therefore, \textbf{H1} was partially supported only for the compliance model in the obesity screening task.

Next, regarding \textbf{H2}, which is related to AI-attention display, there was a negative influence from the AI-attention display on emotional trust for the reliance model in the obesity screening task. Also, there were negative influences from the AI-attention display on cognitive and emotional trust for the compliance model in the drowsiness detection task. Therefore, \textbf{H2} was supported for the reliance and compliance models in the obesity screening task and for the reliance model in the drowsiness detection task.

Finally, regarding \textbf{H3}, which is related to AI-attention interpretability, there was a positive influence from the interpretability of AI attention on emotional trust for the reliance model in the obesity screening task. Also, there were positive influences from the interpretability of AI attention on cognitive and emotional trust for the compliance model in the drowsiness detection task. Therefore, \textbf{H3} was only supported for the compliance model and partially supported for the reliance model in the drowsiness detection task.

\section{General Discussion}

First of all, the obesity screening was easier than the drowsiness detection for the humans as in Experiment 1. Therefore, AI reliability was considered to be perceived more easily in the obesity screening than in the drowsiness detection. Consistent results were observed in SEM. On the basis of this assumption, we discuss the results.

Regarding \textbf{H1}, in the compliance procedure, the participants could observe all the answers of the AI, but in the reliance procedure, they could not when they answered by themselves. Since people tend to develop positive attitudes toward AI when accessing information on AI performance is easy~\cite{Kayande2009how}, in a situation where the task is easy, as in the case of the obesity screening, and the AI performance is always observable, people might positively accept the purpose. Contrarily, in a situation where the task is less easy, as in the case of drowsiness detection, and the AI performance is not always observable, the participants might develop a negative attitude toward the AI and have doubts about the purpose or feel that it is uninterpretable.  

Moreover, regarding \textbf{H2}, there were negative effects from displaying AI attention in the drowsiness detection task. People have a tendency to look for justification when they receive answers or decisions from AI~\cite{Lu2017justifying}. However, in the obesity screening task, the participants might not have needed to pay very much attention to the heat maps as justification because the reliability of the AI could be easily perceived. In comparison, in the drowsiness detection task, since the reliability of the AI seemed more difficult to perceive, participants were considered to pay more attention to the heat maps as justification. However, there were heat maps with low interpretability in accordance with the experimental conditions, and thus, the participants who saw the low-interpretability heat maps might have greatly decreased trust. The same phenomenon was also found in a previous study where people distrusted the AI when they did not perceive the rationale of the AI~\cite{Goodwin2013}. 

Furthermore, regarding \textbf{H3}, as above, in the obese screening task, the participants were considered to not pay very much attention to the heat maps. Therefore, there was no effect found from the attention interpretability. However, in the drowsiness detection task, the attention interpretability especially increased emotional trust. Emotional trust is known to be increased by anthropomorphism and human-like behaviors~\cite{Glikson2020human}. There is a possibility that the heat maps with high interpretability in the drowsiness detection task might have made participants feel that the AI had human-like behaviors and increased their emotional trust.

Finally, there were negative influences from cognitive trust on reliance and compliance in the drowsiness detection task. In this study, the task AI was explained as learning from the participants' answers and becoming smarter. There is a possibility that the participants who had higher cognitive trust in the drowsiness detection, a high difficulty task, might have tended to have the partner AI learn more to become smarter.

\section{Conclusion}

This study investigated how explanations of AI affect human trust in and use of AI, using XAI, which shows its purposes and attention as explanations. We generated high- and low-interpretability heat maps showing AI attention, conducted Experiment 1 to confirm their interpretability, and performed Experiment 2 to investigate the effects of the purposes and attention of AI. SEM analyses revealed that (1) displaying the purpose of AI positively influenced trust when the participants complied with AI in an easier task and negatively influenced trust when they relied on AI in a more difficult task, (2) just displaying the heat maps negatively influenced trust when the participants in a more difficult task, and (3) the heat maps positively influenced trust according to their interpretability in a more difficult task.


\bibliographystyle{apacite}
\setlength{\bibleftmargin}{.125in}
\setlength{\bibindent}{-\bibleftmargin}
\bibliography{cogsci23-m}

\end{document}